\pdfoutput=1
\documentclass[aps,prl,twocolumn,showpacs,superscriptaddress]{revtex4-1}
\usepackage{epsfig,amsopn,amssymb,amsmath,amsfonts}
\usepackage{graphicx}
\usepackage{color}
\usepackage{wasysym}

\newcommand{\rr}{\mathbf{r}}
\newcommand{\s}{\mathbf{s}}
\newcommand{\ttt}{\mathbf{t}}
\newcommand{\e}{\mathbf{e}}
\newcommand{\0}{\mathbf{0}}

\begin{document}

\title{Long-range correlations in a locally driven exclusion process}
\author{Tridib Sadhu}
\affiliation{Institut de Physique Théorique, CEA/Saclay, Gif-sur-Yvette Cedex, France.}
\author{Satya N. Majumdar}
\affiliation{Univ. Paris-Sud, CNRS, LPTMS, UMR 8626, Orsay F-01405,
France.}
\author{David Mukamel}
\affiliation{Physics of Complex Systems, Weizmann Institute of Science, Rehovot
7610001, Israel.}
\date{\today}
\begin{abstract}
	We show that the presence of a driven bond in an otherwise diffusive
	lattice gas with simple exclusion interaction results in long-range
	density-density correlation
    in its stationary state. In dimensions $d>1$ we show that in the
    thermodynamic limit this correlation
    decays as $C(\rr,\s)\sim (r^2+s^2)^{-d}$ at large distances $r$ and $s$ away from the
    drive with $|\rr-\s|>>1$. This is derived using an electrostatic analogy
    whereby $C(\rr,\s)$
    is expressed as the potential due to a configuration of electrostatic charges distributed
    in $2d$-dimension. At bulk density $\rho=1/2$ we show that the potential is
    that of a localized quadrupolar charge. At other densities the same is
    correct in leading order in the strength of the drive and is argued
    numerically to be
    valid at higher orders.
\end{abstract}
\pacs{05.40.-a, 05.70.Ln, 05.40.Fb}
\maketitle
Unlike systems in equilibrium, non-equilibrium stationary states often
exhibit generic scale invariance, \textit{i.e.}, spatial correlation of local
thermodynamic variables has power-law tail
\cite{BTW,dorfman,bertini,spohn,garrido,GRINSTEIN}.
This intriguing feature has been demonstrated in numerous models of non-equilibrium
systems where an external drive prevents the system from achieving thermodynamic
equilibrium.
In particular, for a diffusive system coupled to two reservoirs at unequal
densities the correlation of density fluctuations has been found to decay as
$1/r^{d-2}$ in dimensions $d>2$ \cite{spohn}. On the other hand in the presence
of uniform bulk drive the correlation decays as $1/r^{d}$ \cite{garrido}. The role played
by the drive-induced anisotropy in generating power-law correlation in systems with
conserving dynamics has been elucidated in \cite{GRINSTEIN}.
In some cases, long-range correlations can be related to the non-local nature of the
large-deviation function of the density profile \cite{bernard3,bernard1}.

A natural question to ask is what happens when the drive is spatially localized in the bulk. Such
a drive can be thought of as a local perturbation that breaks detailed balance.
Moreover, the drive breaks translational symmetry. This is a crucial difference
from most of the models studied earlier which has translational symmetry
\cite{garrido,spohn}.

In an earlier work we studied the change in density profile induced by a driving bond in
Symmetric Simple Exclusion Process (SSEP) and in a system of Non-Interacting
(NI) particles \cite{SMM}. In both cases, we found that the density in $d$-dimension
approaches its bulk constant value algebraically as $1/r^{d-1}$ with the distance $r$ away from the
position of the drive. It would be of great interest to compare the steady states of both cases in more detail
by considering the density-density correlation function. In the case of NI particles the
correlation vanishes in the thermodynamic limit. The correlation in SSEP has recently been studied in $d=1$ dimension \cite{bernard2}. It
has been shown that the correlation scales as $1/L$ for a system of length
$L$, and thus it vanishes in the thermodynamic limit. This is related to the
vanishing of the current in the stationary state in the $L\rightarrow\infty$ limit.
The density-density correlation of models of local drive with interacting
particles in higher dimensions has not been considered. As we show below, the
interaction drastically modifies the steady state of the system, generating
long-range correlations which persist in the thermodynamic limit in $d \ge 2$
dimensions.

In this Letter, we study the density-density correlation of SSEP with single
driven bond in $d$-dimensions. We show that due to the fact that the local current
does not vanish in the thermodynamic limit for $d>1$, density-density correlations do
not vanish and they are long ranged. In particular we show that at large
distances $\rr$ and $\s$ away from the drive with $|\rr-\s|>>1$, the density-density
correlation function corresponds to a quadrupolar electrostatic potential in $2d$-dimension
which decays with a power-law tail as
\begin{equation}
	C(\rr,\s)\sim \frac{f(\hat{\rr},\hat{\s})}{(r^2+s^2)^{d}},
\end{equation}
with $f$ being an anisotropic function of the unit directions $\hat{\rr}$,
and $\hat{\s}$.  This is in contrast with the case of
boundary-drive \cite{spohn} where the correlation vanishes in
the thermodynamic limit, in any dimension.

\begin{figure}
	\centering{\includegraphics[scale=.41]{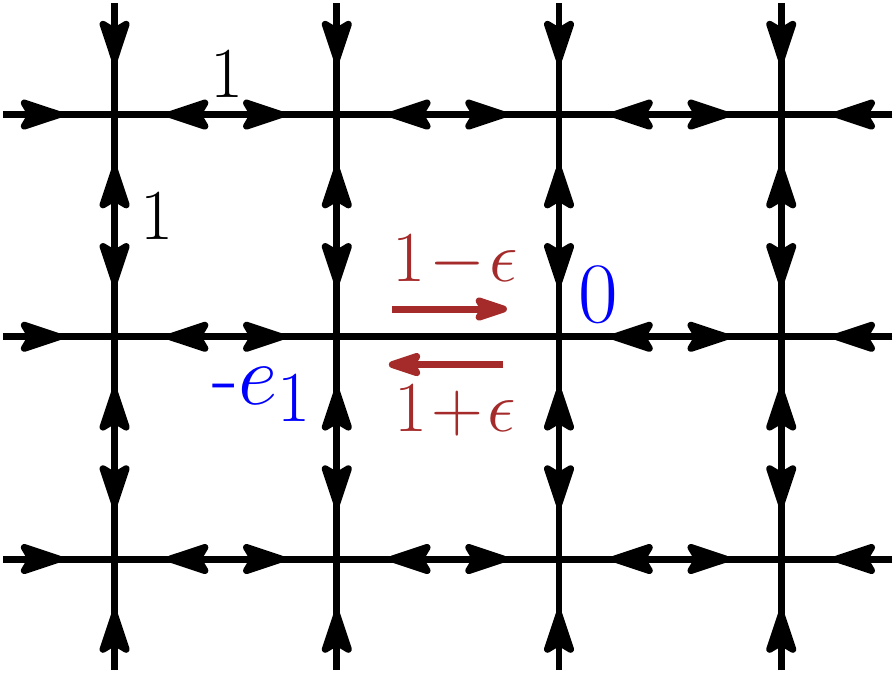}}
\caption{Exclusion process on a periodic square lattice with
asymmetric jump rates between sites $-\e_1$ and $\0$, and symmetric
elsewhere. \label{fig:lattice}}
\end{figure}

The model is defined on a lattice $[-L,L]^d$ in
$d-$dimension with periodic boundary condition. The lattice sites are denoted by
the Cartesian coordinates $\rr=x_1 \e_1+\dots + x_d \e_d$,
where $\e_k$ is the unit vector in the $k$-th direction. The
particles interact with symmetric exclusion
\textit{i.e.} any site is occupied by at most one particle at a time. The
particles hop across bonds with rate $1$ which is symmetric for all bonds
except the one between sites $\rr=\mathbf{0}\equiv\{0,\dots,0\}$ and
$\rr\equiv\mathbf{-\e}_{1}\equiv\{L-1,0,\dots,0\}$. Across this bond the hopping
rate is asymmetric with rate
$(1-\epsilon)$ in the $\e_1$ direction and $(1+\epsilon)$ in the reverse
direction (see Fig. \ref{fig:lattice}). The asymmetric rate drives a circulating particle current around the
bond, which does not vanish in the $L \rightarrow\infty$ limit.

We start by considering the steady state density profile. Let $n(\rr)$ denote the occupation number of site $\rr$ which takes values $1$ or $0$,
depending on whether or not the site is occupied. The density in the stationary
state is defined by
$\phi(\mathbf{r})=\langle n(\mathbf{r})\rangle$, where the angular brackets
denote ensemble average. It is straightforward to show that in the stationary
state $\phi(\mathbf{r})$ satisfies \cite{SMM},
\begin{equation}
	\Delta\phi(\mathbf{r})=\epsilon \langle Q \rangle
\left(\delta_{\mathbf{r},\mathbf{0}}-\delta_{\mathbf{r}+\mathbf{e}_{1},\0} \right),
\end{equation}
where
\begin{equation}
	Q= n(-\mathbf{e}_{1})\left[1- n(\mathbf{0})\right] +
n(\mathbf{0})\left[1- n(\mathbf{-e}_{1})\right],
\label{eq:Q}
\end{equation}
and $\Delta$ denotes Laplacian operator on the lattice.
In an electrostatic analogy $\phi(\rr)$ corresponds to the potential due to a dipole
across the driven bond, and at large distances it decays to its bulk constant
value as $1/r^{d-1}$ with the distance $\rr$ away from the
position of the drive.

We use similar electrostatic analogy to determine the spatial correlation of the
density fluctuations defined as $C(\rr,\s)=\langle
n(\rr)n(\s)\rangle -\phi(\rr)\phi(\s)$. Using the
identity $n(\rr)^{2}=n(\rr)$ the correlation at $\rr=\s$
can be expressed in terms of the density as $C(\rr,\rr)=\phi(\rr)\left[
1-\phi(\rr)
\right]$. It is thus convenient to subtract this value and define
\begin{equation}
	c(\rr,\s)=C(\rr,\s)-\phi(\rr)\left[ 1-\phi(\rr) \right]\delta_{\rr,\s}.
\end{equation}
Thus $c(\rr,\rr)=0$ for any $\rr$ and $c(\rr,\s)$ vanishes at large distances,
$\rr$ and $\s$.

From the dynamics of the model it can be shown
(see Supplemental Material \cite{sup}) that in the stationary state, $c(\rr,\s)$ follows
a Poisson equation on a $2d$-dimensional lattice made of $(\rr,\s)$ vectors.
\begin{equation}
\Delta c(\rr,\s)=\sigma(\rr,\s),
\label{eq:Poisson}
\end{equation}
with $\sigma(\rr,\s)$ given below in \eqref{eq:sigma}-\eqref{eq:sigma3}.
In an electrostatic analogy, $c(\rr,\s)$ is the potential due to the charge
$\sigma(\rr,\s)$ on the $2d$-dimensional lattice.
The large distance profile of $c(\rr,\s)$ is determined by the lowest
non-vanishing multi-pole moment of the charge $\sigma(\rr,\s)$ which we show to
be an effectively localized quadrupole near the origin in the thermodynamic limit.

For convenience the charge density
is expressed below as a combination of three charges
\begin{equation}
\sigma(\rr,\s)=\sigma_{1}(\rr,\s)+\sigma_{2}(\rr,\s)+\sigma_{3}(\rr,\s),
\label{eq:sigma}
\end{equation}
where
\begin{eqnarray}
\sigma_{1}\left( \rr,\s \right)&=&
\sum_{\nu=1}^{d}\left(\delta_{\rr+\e_{\nu},\s}+\delta_{\rr,\s}\right)\left[
c(\rr+\e_{\nu},\s)-2c(\rr,\s)\right.\nonumber\\
&&\qquad\qquad\qquad\left.+c(\rr,\s-\e_{\nu})\right]+\left( \rr \leftrightarrow \s \right)
	\label{eq:sigma1}
\end{eqnarray}
\begin{eqnarray}
	\sigma_{2}\left( \rr,\s
	 \right)&=&\sum_{\nu=1}^{d}\left[\left(\phi(\rr+\e_{\nu})-\phi(\rr)\right)^{2}-\delta_{\nu,1}\delta_{\rr,-\e_{1}}\frac{\epsilon^2\langle
	Q \rangle^{2}}{d}
	\right]\nonumber\\
&&\qquad\qquad\qquad\delta_{\s,\rr+\e_{\nu}}+\left( \rr \leftrightarrow \s \right)
	\label{eq:sigma2}
\end{eqnarray}
\begin{eqnarray}
\sigma_{3}\left( \rr,\s \right)&=&\epsilon \langle Q
\widehat{n}(\s)\rangle\left(\delta_{\rr,0}-\delta_{\rr,-\e_{1}}\right)\left(
1-\delta_{\s,\0}-\delta_{\s,-\e_{1}} \right)\nonumber\\
&&\qquad\qquad\qquad+\left( \rr \leftrightarrow \s \right).
	\label{eq:sigma3}
\end{eqnarray}
Here $\widehat{n}\left( \rr \right) \equiv n(\rr)-\phi\left( \rr \right)$ and
$(\rr\leftrightarrow \s)$ denotes the term obtained by interchanging
$\rr$ and $\s$ in the preceding expression. The latter is related to the symmetry of $c(\rr,\s)$ under exchange of
$\rr$ and $\s$.

The charge $\sigma_{1}(\rr,\s)$ is a linear function of the the correlation itself and is
introduced to include the $\rr=\s$ sites in the Poisson equation (see Supplemental Material \cite{sup}). The other two
charge densities result from the asymmetric hopping rates across the driven
bond. Both vanish for $\epsilon=0$ which corresponds to equilibrium state where
$c(\rr,\s)=0$ is the solution in the thermodynamic limit.

Before analyzing the charge distribution in detail it is instructive to examine its symmetry.
The model is invariant under two symmetry operations. The first is a combination of
reversal of the drive ($\epsilon$) and space inversion,
\begin{equation}
	c_{-\epsilon,\rho}(-\rr-\e_{1},-\s-\e_{1})=c_{\epsilon,\rho}(\rr,\s).
	\label{eq:symm1}
\end{equation}
The second is a combination of space inversion and particle-hole exchange
\begin{equation}
	c_{\epsilon,1-\rho}(-\rr-\e_{1},-\s-\e_{1})=c_{\epsilon,\rho}(\rr,\s).
	\label{eq:symm2}
\end{equation}

The symmetry of the model enables one to extract the large distance profile of the correlation function
for $\rho =1/2$ without actually solving the Poisson equation.
At this density the second symmetry \eqref{eq:symm2} implies that $c(\rr,\s)$ is
even under space inversion.
Then the large distance profile is determined by the lowest non-vanishing
``even'' multi-pole moment which could be monopole, quadrupole
\textit{etc}.
Since, as shown below, the net charge vanishes at all densities, and the quadrupolar
component is shown to be non-vanishing and effectively localized near the origin, the correlation function for $\rho=1/2$ decays at
large distance as the electrostatic potential of a quadrupole moment in $2d$
dimension, namely, $c(\rr,\s)\sim (r^2+s^2)^{-d}$.

The fact that the monopole moment of the charge distribution vanishes in the case of periodic
boundary conditions is evident from integrating the Poisson equation \eqref{eq:Poisson}. It can be
shown, by using the dipolar form of the density $\phi(\rr)$, that the same is true for an infinite
lattice with boundary condition of vanishing correlation at infinity.

In order to analyze the correlation function at densities $\rho \ne 1/2$ we examine in more detail
the expressions for the charge density. While the charges $\sigma_{1}$ and
$\sigma_{2}$ are functions of the density $\phi$ and the correlation function
$c(\rr,\s)$ itself, the $\sigma_{3}$ charge is a function of higher order
correlations.
To see this let us examine the expression for $\langle Q
\widehat{n}(\rr)\rangle$ in $\sigma_{3}$ which using \eqref{eq:Q} yields
\begin{eqnarray}
	\langle Q \widehat{n}\left( \rr \right)\rangle=&&\left(
	1-2\phi(\0)\right)c(\rr,-\e_{1})+\left(
	1-2\phi(-\e_{1})\right)c(\rr,\0)\nonumber\\
&&	-2c(\rr,\0,-\e_{1}),
\label{eq:Qn}
\end{eqnarray}
for $\rr\ne\0$, $-\e_{1}$. Here $c(\rr,\0,-\e_{1})$ is the three point correlation function which
itself depends on higher order correlation functions. This
hierchical dependence of correlation makes the solution difficult.
However, the hierarchy can be handled by a perturbative expansion in $\epsilon$, and in
principle one could solve the problem order by order in $\epsilon$. This becomes possible due to the
pre-factor $\epsilon$ in the expression of $\sigma_{3}(\rr,\s)$ in
\eqref{eq:sigma3}.

Consider an expansion around the zero drive state as
\begin{equation}
	c(\rr,\s)=c_{0}(\rr,\s)+\epsilon c_{1}(\rr,\s)+\epsilon^{2}
	c_{2}(\rr,\s)+\epsilon^{3}c_{3}(\rr,\s)+\dots
	\label{eq:expansion}
\end{equation}
The  first term, $c_{0}(\rr,\s)$ corresponds to the correlation in equilibrium
which is zero in the thermodynamic limit. Also, $\langle Q \widehat{n}(\rr)
\rangle$ in \eqref{eq:Qn} vanishes in
equilibrium. Using these observations in equations
\eqref{eq:sigma}-\eqref{eq:sigma3} it is evident
that $c_{1}(\rr,\s)=0$ in the thermodynamic limit.
Then, the leading non-vanishing term in \eqref{eq:expansion} is of order
$\epsilon^{2}$.

A similar argument can be used to show that the three point correlation $c(\rr,\s,\ttt)$ is
of order $\epsilon^{3}$ or higher. An immediate consequence of the latter is that, in the thermodynamic limit, the terms up to order $\epsilon^{3}$
in \eqref{eq:expansion} do not depend on the higher order correlations, and can
in principle be determined using equations
\eqref{eq:sigma}-\eqref{eq:sigma3}.

We proceed by noting that the first symmetry relation in \eqref{eq:symm1} when applied to \eqref{eq:expansion} implies that all
the  even terms in $\epsilon$ are invariant under space inversion,
whereas the odd terms change sign. Thus,
\begin{eqnarray}
	c_{2k}(-\rr-\e_{1},-\s-\e_{1})&=&c_{2k}(\rr,\s)\\
	c_{2k+1}(-\rr-\e_{1},-\s-\e_{1})&=&-c_{2k+1}(\rr,\s),
\end{eqnarray}
with integer $k\ge0$.
Thus, all the even terms in $\epsilon$ are generated by
charges of ``even'' multi-pole moments like monopole, quadrupole \textit{etc}.
As argued earlier, the monopole vanishes at all densities. 
Thus, to leading order in $\epsilon$ the large distance correlation function is
expected to be determined by the quadrupole component of the charge at any
density.

\begin{figure}[t]
\centering{\includegraphics[width=0.48\textwidth]{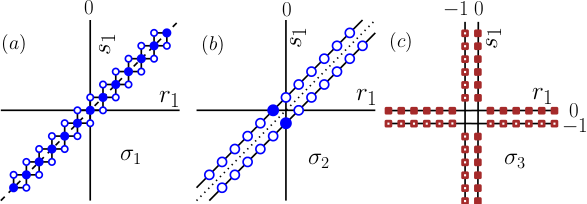}}
\caption{The distribution of charges on the $(r_{1}\e_{1},s_{1}\e_{1})$ plane.
The charge ($a$) $\sigma_{1}$ ($b$)
$\sigma_{2}$ and ($c$) $\sigma_{3}$.\label{fig:charge distribution}}
\end{figure}

In order to demonstrate that the correlation function indeed decays at large
distances as $1/(r^2+s^2)^d$ we analyze the charge in more detail arguing that its
quadrupolar component does not vanish and that this component is effectively
localized near the origin.
A schematic representation of the charge distribution
in the $(r_{1}\e_{1},s_{1}\e_{1})$ plane
is shown in Fig.\ref{fig:charge distribution}. The charges are distributed on
$d$-dimensional planes whose
intersection with the $(r_{1}\e_{1},s_{1}\e_{1})$ plane are straight lines.

The simplest charge to analyze is $\sigma_{1}$ in \eqref{eq:sigma1}
which using $c(\rr,\rr)=0$ yields
\begin{eqnarray}
	\sigma_{1}(\rr,\s)&=&\sum_{\nu}\left[
	\left(2c(\rr+\e_{\nu},\rr)+2c(\rr-\e_{\nu},\rr)\right)\delta_{\s,\rr}\right.\nonumber\\
	&&\left.-2c(\rr,\s)\left(
	\delta_{\s,\rr+\e_{\nu}}+\delta_{\s,\rr-\e_{\nu}}
	\right)\right],
\end{eqnarray}
From this expression it is clear that the charge
is a collection of quadrupoles along $\rr=\s$ plane, each made of four charges
$2c(\rr+\e_{\nu},\rr)$ of alternating signs placed on the corners of the square
plaquettes shown in Fig.\ref{fig:charge distribution}$a$. The filled and empty
circles denote charges of opposite sign.

The $\sigma_{2}$ charge, shown in Fig.\ref{fig:charge
distribution}$b$, is made of positive
charges ($\Circle$) on $\s=\rr\pm\e_{\nu}$ planes and a localized
negative charge ($\CIRCLE$) at $(-\e_{1},\0)$ and $(\0,-\e_{1})$. Using
the density profile $\phi(\rr)$ it is straightforward to show that the quadrupolar moment is non-zero.

The $\sigma_{3}$ charge is shown in Fig.\ref{fig:charge distribution}($c$) where
($\square$) denotes a charge $\langle
Q\widehat{n}(\rr)\rangle$, while ($\blacksquare$) denotes a charge $-\langle
Q\widehat{n}(\rr)\rangle$.
At $\rho=1/2$ one has
$\langle Q \widehat{n}(-\rr-\e_{1})\rangle=-\langle
Q\widehat{n}(\rr)\rangle$. This can be shown using \eqref{eq:Qn} and
noting that the three point correlation is odd under space inversion for this
density. Then clearly the quadrupole moment is non-zero.
At other densities, using continuity with variation of $\rho$, the quadrupole
moment remains non-zero.

In order for the large distance behavior of the correlation function to be that
of a quadrupole, one has to demonstrate that the quadrupole moment of the charges
$\sigma_1$, $\sigma_2$ and $\sigma_3$ is effectively localized around the
origin. By this we mean that the quadrupolar moment density decays sufficiently
fast from the origin such that the large distance decay of the potential behaves
as that of a localized quadrupole.
This can be argued using the following observation: consider a potential in
$2d$-dimension due to distributed quadrupoles on a $d$-dimensional plane with
moment density $\gamma(\mathbf{R})$
at $\mathbf{R}$. For $\gamma(\mathbf{R})$ which decays at large distances faster
than $R^{-d}$, the generated potential can be shown to have
the same power-law tail as the potential due to a localized quadrupole at origin
(see Supplemental Material \cite{sup}).
Indeed one can show that the quadrupole density of each of the three charge
densities decays at large $R$ faster than $R^{-d}$. The charge density
$\sigma_{1}(\mathbf{R})$ is proportional to the reduced correlation
$c(\mathbf{R})$. Therefore a quadrupolar profile of $c(\mathbf{R})$ implies that
$\sigma_{1}(\mathbf{R})$ decays as $1/R^{2d}$ which is faster than required for
effective localization, indicating self-consistency of the solution. Similar
argument also applies for $\sigma_{3}(\rr,\s)$ whose strength $\langle Q
\widehat{n}(\rr)\rangle$ decays as $c(\rr,\s)$. The charge
$\sigma_{2}(\mathbf{R})$ is composed of two parts: a distributed positive charge
$\vert\nabla \phi(R)\vert^{2}$ which decays as
$R^{-2d}$, and a localized negative charge at the origin. These two charges effectively generate
localized quadrupole moment. We thus conclude that the three charges generate
correlation function $c(\rr,\s)$ which at large distance decays as the potential
generated by a localized quadrupole in $2d$ dimensions.

We now consider the effect of the odd terms in the expansion
\eqref{eq:expansion}. At density $\rho=1/2$ all odd terms in the expansion
vanish due to inversion symmetry and the decay of correlation is expected to be
that of quadrupole charge for any $\epsilon$. However, for densities $\rho\ne
1/2$ the odd terms do not vanish and they can support odd multi-pole moments such
as dipole, hexapole \textit{etc}. We have shown above that in the thermodynamic
limit $c_{1}(\rr,\s)$ vanishes. Thus to leading order in $\epsilon$ (namely
$\mathcal{O}(\epsilon^{2})$) the
density-density correlation function decays as the potential of a localized quadrupole.
Our numerical studies in $d=2$ dimensions,
presented below, show that the decay of correlation indeed corresponds to that
of a quadrupolar charge in $d=4$ dimensions even at finite $\epsilon$. This suggests that the dipole
contribution which can come from higher order odd terms, such as
$c_{3}(\rr,\s)$, vanish in thermodynamic limit.

In order to test our result we calculated the
correlation function in $d=2$ dimensions numerically, and show that at large
distance it indeed decays as an electrostatic potential generated by a
quadrupole at origin in $d=4$ dimensions.
The difficulty in determining $c(\rr,\s)$ has to do with the charge $\sigma_{3}$ in
\eqref{eq:sigma3} which depends on the three
point correlation. Ignoring the three point
correlation, which is justified for small $\epsilon$, we determine $c(\rr,\s)$ by numerically solving the 
Poisson equation \eqref{eq:Poisson}.

The solution is obtained by iterating the equation
\begin{equation}
	\partial_{t}c(\rr,\s)=\Delta c(\rr,\s)-\sigma(\rr,\s),
\end{equation}
on a four dimensional integer lattice $[-40:40]^{4}$. The long time profile
$c(\rr,\s)$ is the solution of the \eqref{eq:Poisson} and corresponds to the
correlation for a two-dimensional system. For a fast convergence we have
conditioned $c(\rr,\s)=0$ at the boundary. As the charges are
themselves a function of $c(\rr,\s)$, at every time step they are updated with
the existing profile. The profile after $10^6$ iterations varies by 
$\mathcal{O}(10^{-10})$ and considered as the asymptotic solution.

\begin{figure}
	\centering{\includegraphics[width=0.4\textwidth]{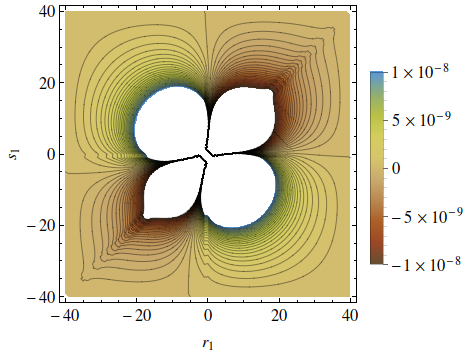}
	\caption{Contour plot on $(r_{1}\e_{1},s_{1}\e_{1})$ plane of the numerical solution of the Poisson equation
	\eqref{eq:Poisson} on a $4$-dimensional lattice. (color online)
	\label{fig:contour}}}
\end{figure}
\begin{figure}
	\centering{\includegraphics[width=0.40\textwidth]{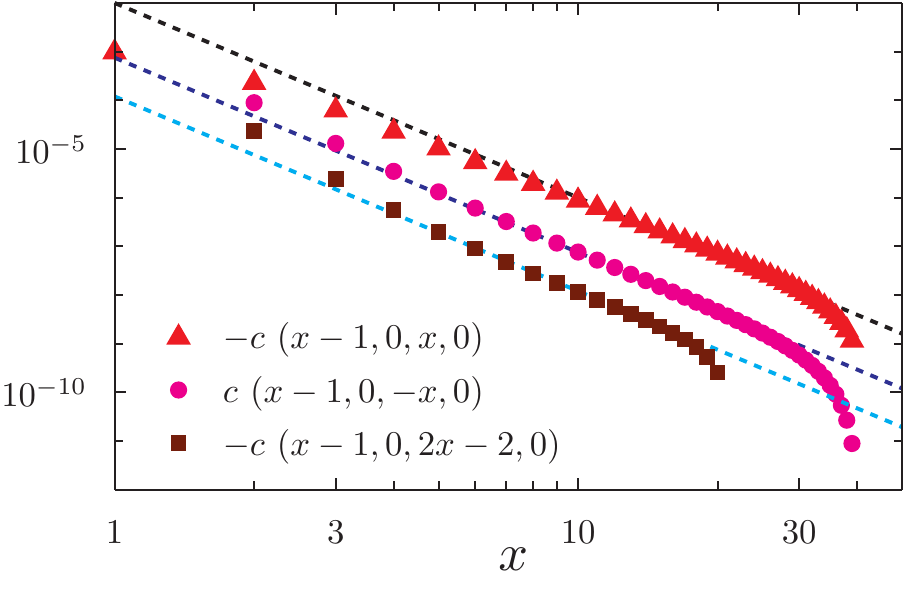}
	\caption{Log-Log plot of the $c(\rr,\s)$ along three different directions. The dashed lines
	denote a power-law fit by $1/x^{4}$. The ($\blacktriangle$) data is
	shifted by one unit in log scale for better presentation. \label{fig:numerical}}}
\end{figure}

The resulting profile on the $(r_{1},0,s_{1},0)$ plane for $\epsilon=1/2$ and $\rho=1/4$ is shown in
Fig.\ref{fig:contour}. The different colors denote the value of $c(\rr,\s)$ as
indicated in the color bar. The profile has a slightly broken quadrupolar symmetry at
large distances. The small asymmetry is likely due to finite system
size and higher odd multi-pole moments. In the thermodynamic limit the
asymmetry is expected to vanish at large distances.

The power-law tail of the $c(\rr,\s)$ along three different directions away from
the origin is shown in Fig.\ref{fig:numerical}. The result is consistent with a quadrupolar decay
$1/(r^2+s^2)^{2}$. The exponential decay near the tail is due to finite-size
of the lattice.

To conclude, we have shown that in the presence of a single driven bond in a bulk SSEP
in $d$-dimension, the density-density correlation is long-ranged. Using symmetry
considerations we showed that, for $\rho=1/2$, the correlation at large distances
is expressed in terms of the potential due to a localized quadrupole at origin. At other densities, we
use a perturbative analysis to argue that the same is true in the small drive
($\epsilon$) limit. In particular, we showed that, in the thermodynamic limit, the leading
non-vanishing term of $c(\rr,\s)$ is of order $\epsilon^2$ and is generated
by a localized quadrupole. For general $\epsilon$, the evidence for
the quadrupolar profile of $c(\rr,\s)$ comes from numerical solution of the
stationary equation \eqref{eq:Poisson} for $d=2$ which shows that
$c(\rr,\s)\sim(r^2+s^2)^{-d}$ for large $\rr$ and $\s$.
It would be interesting to see how this quadrupolar decay in correlations may
change in presence of more than one driving bonds.

We thank A. Bar, O. Cohen, B. Derrida, O. Hirschberg and K. Mallick for fruitful
discussions. The support of the Israel Science Foundation (ISF) and the Minerva Foundation
with funding from the Federal Ministry of Education and Research is gratefully
acknowledged. TS acknowledges the support of the Weizmann Institute of Science
and IPhT, CEA-SACLAY, where the work has been carried out.

\bibliography{reference}

\end{document}